\documentstyle [12pt]{article}
\newcommand{\be}{\begin{equation}}
\newcommand{\ee}{\end{equation}}
\newcommand{\bdis}{\begin{displaymath}}
\newcommand{\edis}{\end{displaymath}}
\newcommand{\eg}{\varepsilon}
\newcommand{\fs}{|\delta v(r)|}

\title{On the intermittent energy transfer at  viscous scales
in turbulent flows.}
\author{ R. Benzi$^{1}$, L. Biferale$^1$, S. Ciliberto$^2$,
M.V. Struglia$^1$ and R. Tripiccione$^3$}
\begin{document}
\maketitle
\centerline{$^{1}$  Dipartimento di Fisica, Universit\`{a} ``Tor Vergata''}
\centerline{Via della Ricerca Scientifica 1, I-00133, Rome, Italy.}

\centerline{$^{2}$
      Laboratoire de Physique,  Ecole Normale Sup\'erieure de Lyon }
     \centerline{ CNRS URA1325 - 46 All\'ee d'Italie, 69364 Lyon, France}
\centerline{$^{3}$  INFN, Sezione di Pisa,
S. Piero a Grado, 50100 Pisa, Italy}
\date{ }
\medskip
\centerline{\it Submitted to Europhys. Lett.}

\begin{abstract}
In this letter we present numerical and experimental results
on the scaling properties of velocity turbulent fields
in the range of scales where viscous effects are acting.

A  generalized version of   Extended Self Similarity
capable of describing scaling laws of the velocity structure
 functions down to the smallest resolvable scales is introduced.
 Our findings suggest the absence of any sharp viscous cutoff in
 the intermittent transfer of energy.
 \end{abstract}

\newpage

The word anomalous scaling (AS) usually refers to scaling laws in a physical
system which cannot be deduced from naive dimensional arguments.
It is always
a challenging problem in physics to understand the origin of anomalous
scaling and to formulate a predictive theoretical framework
to compute the anomalous scaling exponents.

\noindent
Among the many physical systems showing anomalous scaling, fully developed
three dimensional turbulence (FDT) has been
 widely investigated in the last few
years (see \cite{urielbook} for a recent overview of the experimental
and theoretical state of the art).  According to Kolmogorov 1941
 theory \cite{k41}
the small scale statistical properties of FDT obey the relation:
\be
<|\delta v(r)|^p> \sim A_p (\frac{r}{L})^{p/3}v_0^p \sim A_p \eg^{p/3} r^{p/3}
\label{eq:scaling}
\ee
where $\delta v(r) = v(x+r) -v(x)$ is the difference of velocity
at scale $r$, $v_0$ is the  {\it rms} velocity at the integral scale $L$,
$\eg$ is the mean energy dissipation
and the $A_p$'s are dimensionless constants. Eq. (\ref{eq:scaling})
is not satisfied both in real experiments and numerical simulation. Indeed,
one has to replace it by anomalous (also known as intermittent) scaling:
\be
<|\delta v(r)|^p> \sim B_p (\frac{r}{L})^{\zeta(p)}v_0^p ,
\label{eq:an_scaling}
\ee
where $\zeta(p)$ is now a non-linear function of its argument.

\noindent
At variance  with  expression (\ref{eq:scaling}), the scaling
 (\ref{eq:an_scaling})
is anomalous in the sense that it cannot be deduced by naive dimensional
counting. In order to get a more precise measurements of the $\zeta(p)$
exponents and to highlight the anomalous scaling, it has been proposed
in \cite{ess} \cite{ess2}
to look at the self-scaling properties of
the velocity structure functions, namely:
\be
<|\delta v(r)|^p> \sim <|\delta v(r)|^q>^{\beta(p,q)}.
\label{eq:ess}
\ee
This new way of looking at the scaling properties has been tested in
many different experimental and numerical instances \cite{ess_rev}.
In all  cases,
when  small-scales  homogeneity and isotropy were satisfied, a dramatic
improvement in the width of the scaling region was observed.
 This almost universal
property of turbulent flows was then called  Extended-Self-Similarity
(ESS). ESS must interpreted as the signature of some non trivial
universal physics happening at the transition between the inertial
 and viscous scale.
It tells us that, by using the appropriate functional form, scaling
is present also at scales where
in principle viscous effects should already be important.

The aim of this letter is twofold. First we  discuss in more details some
cases where ESS does not work (shear flows and boundary layers).
Second, we  present a generalized version of ESS (G-ESS) which
turns out to be
much more universal and allows us to draw a concrete theoretical
framework of the energy cascade down to the smallest resolvable scale, i.e.
in a region where no anomalous scaling was supposed to be detected.

The physical outcome of our findings is that whatever is the mechanism
responsible for anomalous scaling in FDT, this mechanism is acting
also at extremely small scales and  within experimental
errors no evidence of a cutoff (due to dissipation) is
observed.

We have performed a direct numerical simulation of 3 dimensional
Navier-Stokes eqs. for a
Kolmogorov flow (see \cite{bs} for technical details).
The flow is forced such that the stationary solution has
a non-zero spatial dependent mean velocity
$<\vec{v}(\vec{x})> = \hat{x} sin(\frac{8\pi}{L}z)$,
where $\hat{x}$ is the versor in the direction $x$, and $ L$ is the
integral scale.

\noindent
In figures $1a$ and $1b$ we show the standard ESS analysis by plotting
$<\fs^6>$ versus $<\fs^3>$ for two specific levels $z_a$ and $ z_b$,
where $z_a$
was chosen at  minimum shear and $z_b$ at maximum shear (in this case
$<\cdots>$ must be interpreted as averages over time integration
at fixed z-level). The $Re_{\lambda}$ number of the simulation was $40$  and
no scaling laws were present if examined as a function of the
physical scale $r$.
Nevertheless, it is clear from figure $1a$ that ESS is observed for the case
of minimum shear and it is not observed  for the case of maximum shear
(figure $1b$). Violations
of ESS have already been reported in other cases where strong
shear effects were argued to be relevant \cite{sreenivasan} \cite{ess2}.

In order to understand this phenomenological transition  between
strong shear and weak shear flows let us recall two recent results obtained
by using ESS concepts.

\noindent
The first is a generalized form of the Kolmogorov Refined
Similarity Hypothesis (KRS)
\cite{kolmo_ref}:
\be
<\fs^p> \sim <\epsilon(r)^{p/3}> <\fs^3>^{p/3},
\label{eq:krsh}
\ee
where $<\epsilon(r)^{p/3}>$ is the dissipation energy
averaged over a box of radius $r$. The second result
follows from the moment hierarchy recently proposed
in \cite{sl} and rewritten in terms of structure functions:
\be
\frac{<\fs^{p+1}>}{<\fs^{p}>} =
A_{\infty}(r)^{1-\gamma} \left(\frac{<\fs^{p}>}{<\fs^{p-1}>}\right)^{\gamma}.
\label{eq:sl}
\ee
where
\be
A_{\infty}(r)^{1-\gamma}= A_p \left({ <\fs^{6}> \over
<\fs^{3}>^{1+\gamma^3}}\right)^{(1-\gamma)
\over 3(1-\gamma^3)},
\ee
and $\gamma^3 = 2/3$.
Equation (\ref{eq:krsh}) was proposed in \cite{ess_rev} and
checked systematically
in \cite{journaldephysique1}, while equations (\ref{eq:sl})
 has been discussed in \cite{journaldephysique2}. Let us remark
that (\ref{eq:sl}) can be obtained from the original
She-Leveque hierarchy on the energy
dissipation plus the generalized KRS (\ref{eq:krsh}).

\noindent
The novel point of (\ref{eq:krsh}) and (\ref{eq:sl}) is that they holds
 also for values of $r$
where ESS is no longer satisfied.
In order to highlight the previous comment, let us consider
again the Kolmogorov flow
previously observed. In figure $2a$ and $2b$ we show the result of
 the scaling obtained
by using (\ref{eq:krsh}) at the correspondent $z$-levels of
 fig $1a$ and $1b$ and for
$p=6$. As one can see, the generalized KRS is well satisfied in both
cases although for the
$z_b$  ESS is not observed.

Figures 2, and the results obtained in \cite{ess_rev,journaldephysique2},
 suggest that the concept of ESS could be generalized in such a way to take
into account the scaling relations (\ref{eq:krsh}) and
(\ref{eq:sl}) properly.

For this purpose we introduce the dimensionless structure functions:
\be
G_p(r) = \frac{<\fs^p>}{<\fs^3>^{p/3}}.
\label{eq:dsf}
\ee
According to Kolmogorov theory $G_p(r)$ should be a constant both
in the inertial
range and in the dissipative range, although the two constant are not
necessarily thought to be the same. Because of the presence
of anomalous scaling, $G_p(r)$ are no longer constants in the
inertial range and, by
using (\ref{eq:krsh}) we have:
\be
G_p(r) = <\epsilon(r)^{p/3}>
\label{eq:energy}
\ee
Following the results shown in figs 2 and
in \cite{ess_rev,journaldephysique2} equation (\ref{eq:energy})
is valid for all scales and also in the case where ESS is not verified.
 Therefore, it seems reasonable to
study the self scaling properties of $G_p(r)$ or, equivalently,
the self-scaling
properties of the energy dissipation averaged over a ball of size $r$:
\be
G_p(r) = G_q(r)^{\rho(p,q)},
\label{eq:sess}
\ee
where we have by definition:
\be
\rho(p,q) = \frac{\zeta(p)-p/3 \zeta(3)}{\zeta(q)-q/3 \zeta(3)},
\label{eq:anomalie}
\ee
$\rho(p,q)$ is given by the ratio between deviation from the K41 scaling.
It will play an essential
role in our understanding of energy cascade.
Indeed, it is easy to realise that it is the only
quantity that can stay constant along
all the cascade process: from the integral to the
sub-viscous scales. It is reasonable to imagine that the velocity
field becomes  laminar  in the sub-viscous range,
$<\fs^p> \sim r^p$, still preserving
some intermittent degree parametrized by the ratio between
correction to K41 theory.

Let us give a theoretical argument  behind  relation (\ref{eq:sess}). All
of the proposed cascade-model of turbulence based on
infinitely divisible distribution \cite{urielbook}
lead to a  set of scaling exponents with the following form:
\be
\zeta(p) = ap + b f(p),
\label{eq:zetap}
\ee
where $a,b$ are model-dependent constants and
$f$ is the (model-dependent) non-linear function giving
the intermittent corrections to K41. In the   Log-Poisson
model \cite{sl}, for example, we have
$f(p) = 1-(\frac{2}{3})^{p/3}$.
ESS is a statement  concerning the ratio
$\zeta(p)/\zeta(3)$ as a function of the analysed range
of scales. Based on the expression (\ref{eq:zetap}) a possible interpretation
is that the free parameters $a$ and $b$ acquires a
weak scale-dependency near the viscous
cutoff such that the ratio $a/b$ stays constant.
The ESS violation
would mean that for very small $r$ (comparable
with the viscous cutoff) and/or in presence
of a strong shear the $r$-dependency of $a$ and
$b$ is no longer the same. However, if we make
the (very strong) assumption that the non linear transfer of energy is
always working, regardless on
how $a$ and $b$ depend on $r$, then we should
expect that the non linear (anomalous) contribution
to $\zeta(p)$ is independent of $r$, i.e.
$f(p)$ in expression (\ref{eq:zetap})
must not depend on the scale even near and inside the viscous range.
 It follows that the ratio between anomalies (\ref{eq:anomalie}),
using (\ref{eq:zetap})
\[
\rho(p,q)=\frac{f(p)-\frac{p}{3}f(3)}{f(q)-\frac{q}{3}f(3)},
\]
must be a $r$-independent quantity. Thus the scaling relation
(\ref{eq:sess}) should always be observed.

In order to support this claim we have plotted in figures 3
and 4 the scaling of $G_6(r)$ versus $G_5(r)$ for
many different experimental set up\cite{ess,the,abs} done
at different Reynolds number and for some direct numerical
simulation with and without large scale shear.
 As one can see the straight line behaviour
is very well supported. Within experimental
errors (of the order of $3 \%$) no deviations
from the scaling r\'egime are detected.
Similar results are obtained, using different $ G_p(r)$ and $ G_q(r)$.
In a more detailed version of this study \cite{longpaper}
we will discuss  how it is possible to reconcile the idea of multiplicative
cascades with this continuous
energy transfer from the inertial range to the viscous range
and we will present  numerical evidences
that this new scaling behaviour is in disagreement with
the idea of a statistical dependent
viscous cutoff as predicted in all the standard multiplicative
multifractal  models \cite{fv}

Our results may have theoretical and applied
implication. For instance, the presence of intermittent fluctuations
at all scales might cast serious doubts on the validity of
renormalized perturbative expansion of the
NS equations which are usually based on perturbative
expansion around the linearized equations.
\vspace{1.truecm}
\noindent

Discussion with G. Stolovitzky and Stephan Fauve  are kindly acknowledged.
L.B. has been
partially supported by the EEC contract ERBCHBICT941034. R.B.
 has been supported
by the EEC contract CT93-EVSV-0259.

\newpage
FIGURE CAPTIONS:

\noindent
{\bf FIGURE 1a}: Log-Log plot showing ESS scaling for the longitudinal
structure functions, $\fs^6$ versus $\fs^3$. Data are taken from
a direct numerical simulation of a shear flow at $Re_{\lambda}=40$. Each
point corresponds to a space separation of a Kolmogorov scale.
 The computation
of structure functions is performed at points where the shear is minimum.
The dashed line is the best fit for the slope in the
scaling region.

\noindent
{\bf FIGURE 1b}: The same of figure 1a but for points where the shear is
maximum. At variance with the previous case  ESS  is not observed.

\noindent
{\bf FIGURE 2a}: Check of equation (4) for
$p=6$ at  points of minimum shear (in Log-Log scale).
Energy dissipation has been computed by using
 the 1-dimensional surrogate, in order
to compare this result with laboratory experiments
 \cite{ess_rev}. The dashed line
is the best fit.

\noindent
{\bf FIGURE 2b}: The same of figure
2a but for points of maximum shear. Although
in this case ESS is not observed  (see fig. 1b),
generalized-ESS  works within
$3 \%$.

\noindent
{\bf FIGURE 3}:Log-log plot of
$G_6(r)$ versus $G_5(r)$ for different laboratory
and numerical experiments.
Data taken in a wake behind a cylinder,
where standard ESS was not observed
\cite{ess}, (Crosses). Data taken from the
region with log-profile of a boundary
layer (courtesy of G. Ruiz Chavarria) where standard ESS was
not observed (Circles).
Data taken from a direct numerical simulation of
thermal convection \cite{the}
where standard ESS was observed
(Squares). Data from a direct numerical simulation of a channel flow
where standard ESS was not observed \cite{abs} (Triangles).

\noindent
{\bf FIGURE 4}: The same as in figure 3 but for the direct  numerical
simulation of the shear flow.

\end{document}